# The first global-scale 30 m resolution mangrove canopy height map using Shuttle Radar Topography Mission data

Abdullah F. Rahman[1*]

Aslan Aslan[1]

[1]Coastal Studies Lab, University of Texas Rio Grande Valley, 100 Marine Lab Drive, South Padre Island, TX 78597, USA, Phone: 956-761-2644.

[*]Corresponding author: email: abdullah.rahman@utrgv.edu

**Abstract**

High-resolution maps of global mangrove areas have been produced recently using data from Landsat satellites. However, these maps contain only the two-dimensional aspect of mangrove locations, but not the mangrove height information. Mangrove height is an important parameter for quantifying mangrove's biomass and other ecosystem services. No high-resolution canopy height map exists for global mangroves. Here we present the first global mangrove height map at a consistent 30 m pixel resolution derived with digital elevation model (DEM) data from shuttle radar topography mission (SRTM) and refine the current global mangrove area maps by discarding the non-mangrove areas included in the previous maps.

**Introduction**

Mapping mangrove areas of the world is important for research into mangrove functionality, ecosystem dynamics, deforestation and impacts of global climate change etc. All existing maps of global mangoves, including the most recent ones (Giri et al., 2011; Hamilton & Casey, 2016), explore the two dimensional aspect only, i.e., the spatial location of mangroves. The third dimensional aspect of mangroves, i.e., the canopy height, is an essential parameter for understanding the vertical structure, estimating aboveground biomass, and exploring mangrove's ecossytem functions as a habitat for hundreds of species of animals, birds, reptiles, fish, amphibians and insects. Yet, no fine resolution global map of mangrove heights exists currently, although global maps of forest height at coarse resolutions (1 km pixel) have been developed (Simard et al. 2011).

In February of 2000, a synthetic aperture radar (SAR) system flew on board NASA's space shuttle Endeavour during the 11-day STS-99 shuttle radar topography mission (SRTM) and produced a set of quality-controlled digital elevation model (DEM) data on a near-global scale from 56° S to 60° N, at consistent spatial resolutions. SRTM data have been used to estimate canopy heights of forests and mangroves (Fatoyinbo and Simard, 2013; Simard et al., 2006). In theory, SRTM provided digital topographic data above sea level. Since mangroes can grow only in tidal areas at sea level or in very close proximity, the SRTM topographic elevation data of magrove areas can provide a good estiamte of canopy heights. Thus far, only a few areas of the globe have been represented in the studies of mangrove heights using SRTM data, and not all posible heights of mangroves have been included in those studies. Under favorable conditions some species of mangroves can grow up to ~40 m tall in some parts of the world. So, there is a need to include the entire range of 'minimum to maximum' heights of mangrroves for developing a global model of mangrove heights using the SRTM data.

Here we present the first global-scale mangrove canopy height map. We had three objectives in this study. Our first objective was to developed a mangrove height model of SRTM data using globally distributed field-bsaed mangrove height data. Second objective was to use that mangrove model and 30 m resolution SRTM data of global mangrove areas to produce a mangrove canopy height map of global extent, at 30 m pixel resolution. Third was to refine the existing global mangrove area maps by identifyong the areas that had elevation values beyond the range of maximum mangrove heights and by excluding those areas from the current global mangrove maps.

**Results**

We used 450 field-based mangrove canopy heights and their corresponding SRTM values from published and unpublished sources in a scatterplot to develop a linear correlation model:

$$CHM = 0.8955 \times SRTM + 1.81223 \qquad (1)$$

where CHM is the canopy height of a 30 m SRTM pixel. The correlation was linear with a high Pearson's r (r = 0.91) and the scatter was similar for all levels of heights (Figure 1). A k-fold cross-validation analysis of the 'field-data vs. SRTM data' scatter produced average inaccuracy of 2.62 m for MAE and 3.36 m for RMSE.

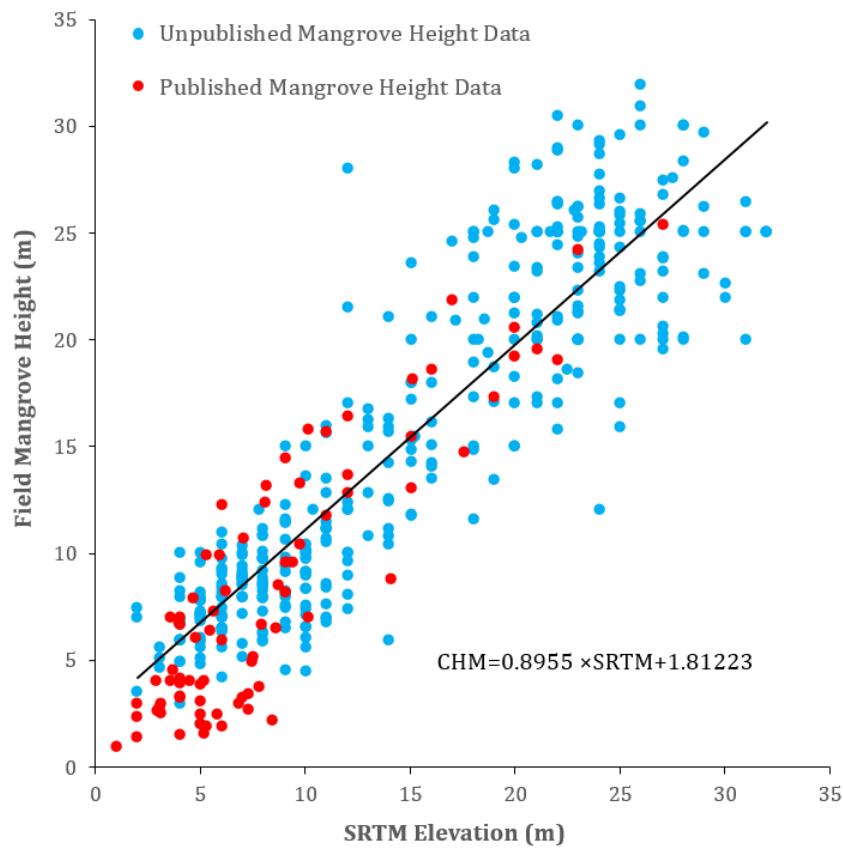

**Figure 1:** Correlation between field-based mangrove height data and STRM data. This correalationwas used to derive the global mangrove height data at 30 m from the SRTM images.

We downloaded the SRTM DEM data from the US Geological Survey National Center for Earth Resources Observation and Science through the Earth Explorer data portal (http://earthexplorer.usgs.gov/). There were 118,784,099 pixels of global mangroves, at 30 m pixel resolution (using the map produced by Hamilton and Casey, 2016). We used Equation 1 with SRTM data of global mangrove areas to estimate the mangrove height at each pixel. The resulting heights ranged from 2.78 to >1,000 m, but the vast majority of the height data were <40 m. Under optimal conditions, mangrove trees can grow to a height of 40 m (Saenger and Snedaker, 1993). Most of the tallest mangrove trees of the world exist in the equatorial areas of Southeast Asia, and they do not grow taller than 40 m. In the Hamilton and Casey (2016) map, 562,800 pixels had height values >40 m. This >40 pixels in the global mangrove map is an error of commission. We excluded the pixels that had >40 m height and created a corrected map of global mangrove areas and canopy heights at 30 m spatial resolution (Figure 2). Converting the remaining pixels to hectares (ha), there were 10,639,916 ha of mangroves globally in 2000 (Figure 2). Our results showed that taller mangroves are present near the equator; and the maximum height of mangroves decline in both hemispheres as the distance from the equator increases.

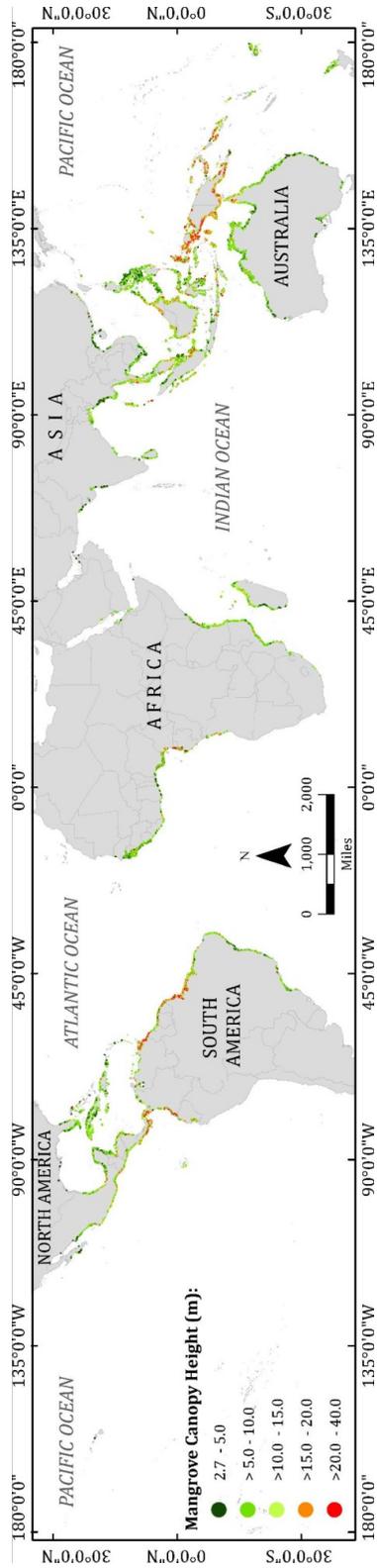

**Figure 2:** Global mangrove height map at 30 m resolution. For display purpose, the map layer is subsampled from 30 m pixels to 0.1º geographic grid.

A positive skew exists in global mangrove heights frequency distribution (Figure 3). The peak of the skew was between 4-9 m heights. Other than the two spikes in the distribution, at 13 m and 22 m, mangrove areas decrease almost exponentially as the heights of mangroves increased.

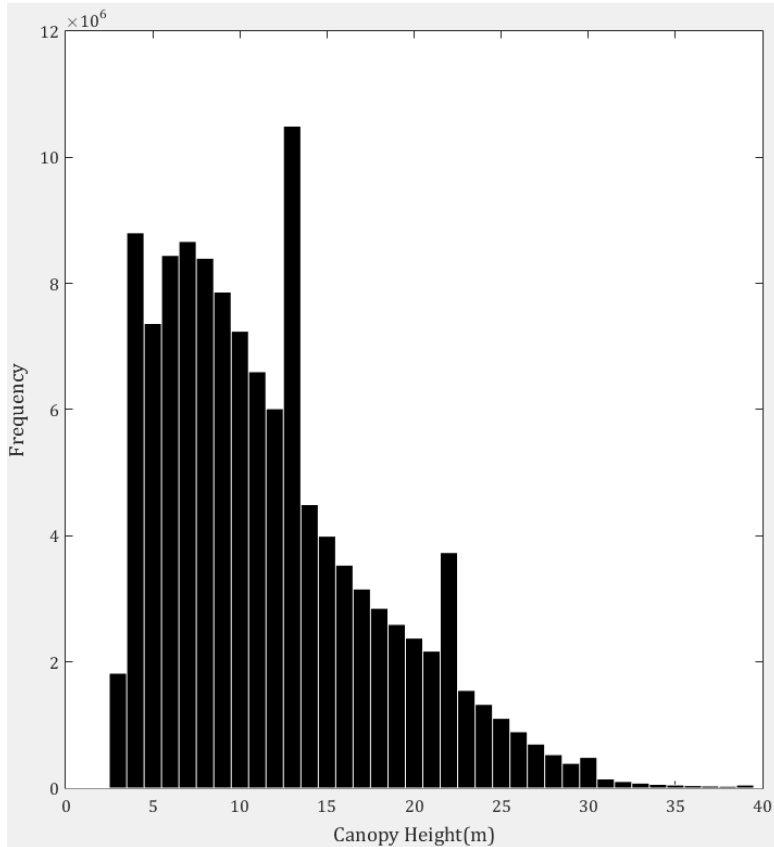

**Figure 3:** Frequency distribution of 30 m mangrove pixels along different canopy heights.

Examples of the distribution of mangroves of different heights in different areas are shown in Figure 4. Coastal areas of Mimica district in the Papua Island of Indonesia show a high concentration of tallest mangroves in the world, reaching a height of 39 m (Figure 4a). The Sundarbans mangrove forest of Bangladesh and India along the northern Bay of Bengal coastline is the world's largest single patch of mangroves, covering more than 1 million ha of land area (Figure 4b). The heights of the Sundarbans mangroves show a southwest to northeast gradient.

Canopy heights are short in the southwest (~2 m), gradually increasing along the northeastern direction, attaining a height of ~20 m. This spatial gradient of canopy height in the Sundarbans mangroves has been recorded in published studies (Rahman et al., 2015). Mangroves of Southern Florida and the Florida Keys are short (~3 m) to medium (~10 m) in height, with the mangroves along the coasts of Ponce de León Bay being the tallest (~20 m) among the entire area (Figure 4c).

(a)

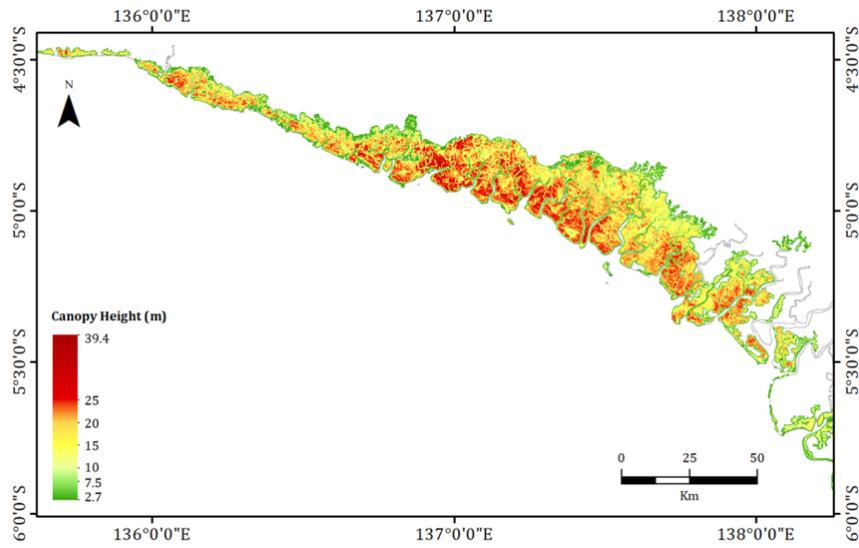

(b)

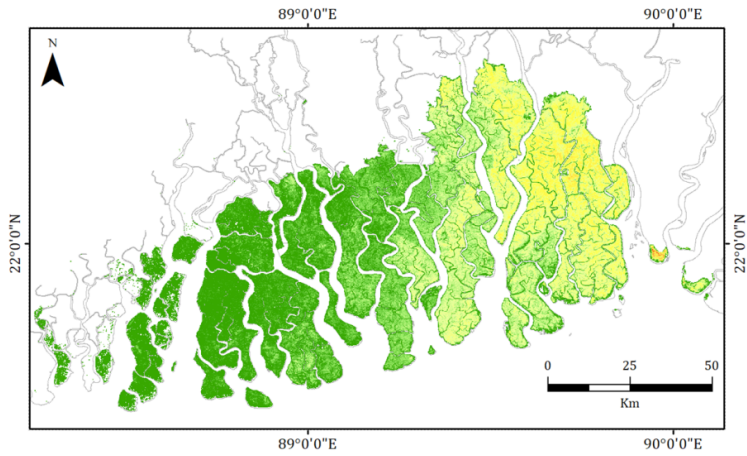

(c)

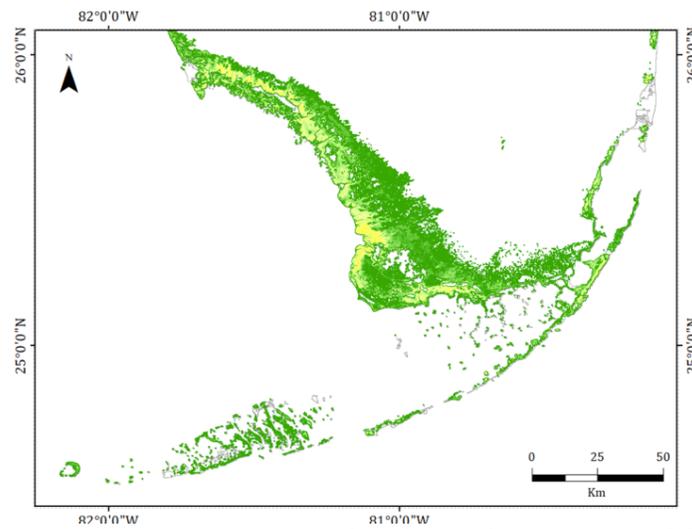

**Figure 4:** Detailed views of mangrove canopy heights at three areas, a) Mmimika, Indonesia, b) Sundarbans, Bangladesh, and c) Florida Keys, USA

In addition to the error of commission in the existing global mangrove maps that include areas that are >40 m tall, our analysis showed that a second level of error is also present in the global 30 m scale mangrove maps (Giri et al, 2011; Hamilton and Casey, 2016). Some coastal areas that have vegetation shorter than 40 m but are not mangroves are also included in the current global mangrove maps. An example of such an area is shown in Figure 6. This area is on the northeastern shore of Bali Island, near Bandjargondol, in the Buleleng Regency of Bali, Indonesia. This area is a valley, was never covered with mangroves, yet its vegetation reflectance values and coastal location led to its inclusion in the global mangrove area maps. Our analysis showed other similar areas exist in different parts of the globe, although the cumulative areas of this type of error seem to be relatively small compared to the areas that were >40 m in canopy height. Our global mangrove height map could not distinguish these areas, and therefore this type of error is still included in the global mangrove height map.

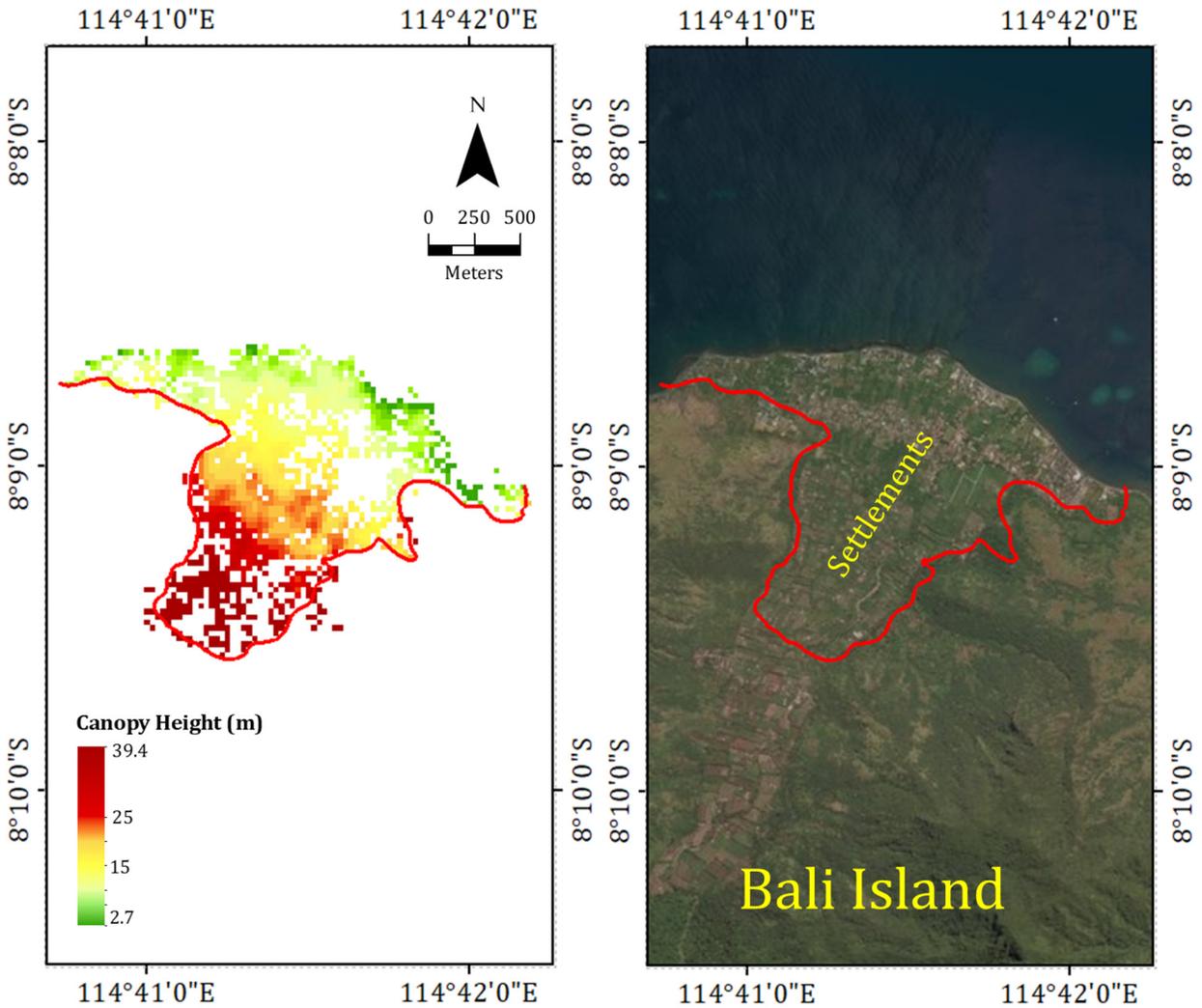

**Figure 5:** An example of the secondary level of error of commission present in current high-resolution global mangrove area maps, where the coastal vegetation that are <40 m tall but are non-mangroves and yet included in the mangrove maps. A Google Earth image of the area is shown on the right and the result of our mangrove height map is shown on the left.

**Discussion**

Previous studies have used SRTM data to derive mangrove heights of limited study areas. Other studies have used synthetic aperture radar (SAR) data from newer sensors, such as European Space Agency's TanDEM-X, to estimate mangrove heights (Lee et al., 2015). Lidar data and stereo images from optical sensors have also been used for this purpose. In this study, we present the first global-scale mangrove height map at 30 m pixel resolution, and correct the global mangrove area map by excluding areas in the existing global mangrove maps that are taller than 40 m. Even though the STRM data sets from 2000 are somewhat dated, recent studies have demonstrated that the average rate of global mangrove deforestation has been <<1% since 2000 (Hamilton and Casey, 2016). Growth and expansion of mangroves is a very slow process. Therefore, the use of SRTM data is considered as appropriate for studying global mangrove heights (Aslan et al., 2016).

Mangrove height map at global scale is an important addition to the tools required for studying and monitoring mangroves. None of the existing global mangrove area maps differentiates among areas under mangroves of different heights. One ha of mangrove under 39 m tall mangroves in Papua, Indonesia, and another ha of mangrove under 3 m tall mangroves in the Florida Keys were basically treated as the same in those two maps – just a ha of mangrove. In reality, taller and shorter mangroves perform fundamentally different ecosystem functions. Taller mangroves provide habitat and food for more species of animals, birds and insects, thus increasing biodiversity. Taller mangroves contain higher above ground biomass compared to the shorter mangroves. Although soil carbon content of mangrove forests depends on many factors, studies across the globe have reported higher carbon content in soils under taller mangroves compared to those under shorter mangroves. Shorter mangrove ecosystems have their own advantages in terms of quick growth and areal expansion, shoreline stabilization and providing spawning and feeding grounds and habitat for fish and other aquatic animals and waterfowls. Therefore, mapping the heights of global mangroves at a consistent spatial scale adds another important dimension to the future studies of mangroves and their ecosystem dynamics.

The global mangrove heights at 30 m pixel resolution shown in Figure 2 correspond well with the results reported by previous mangrove studies: taller mangroves are abundant in the Indo Pacific coastal areas near the equator and shorter mangroves abound in other areas (Saenger and Snedaker, 1993). The frequency distribution of canopy heights demonstrates that majority of global mangroves are <10 m tall, and taller mangroves are rarer (Figure 3). This provides very important information for global mangrove conservation. Given their rarity, conservation of taller mangroves should take precedence. Results of this study would allow the conservation agencies and policy makers to identify the taller mangrove forests in each country and undertake programs to conserve those forests. Our results also show that mangroves of different heights can exist at any particular latitude but the maximum heights of mangroves are limited by their latitudinal locations.

One important aspect of this study was to exclude the areas with >40 m tall canopy heights from the existing maps, and thus further refining the global mangrove map. Our analysis also demonstrated that another type error of commission in current global mangrove maps is the coastal and riparian areas that have <40 m canopy heights and yet they are not mangrove areas (Figure 5). The present study was based on mangrove heights only, and thus was not able to identify the areas with the second type of error or exclude them from the global mangrove map. One suggestion to correct this error would be to superimpose country level residential and land use maps on the mangrove map that we produced in this study, and further refine the mangrove map by excluding the known non-mangrove areas. Use of new higher resolution SAR data, such as from TanDEM-X, may also provide a way to further refine the global mangrove area map by identifying and excluding the constructed areas, housings, etc.